\documentclass[journal]{IEEEtran}
\usepackage{fancyhdr}
\usepackage{amsmath}
\usepackage{subfigure}
\usepackage{soul,cite}
\usepackage{extarrows}
\usepackage{color}
\usepackage{mathrsfs}
\usepackage{graphicx}
\usepackage{booktabs}
\usepackage{algorithm} 
\usepackage{algorithmic} 
\usepackage{amssymb}
\DeclareMathOperator*{\argmax}{arg\,max}
\DeclareMathOperator*{\argmin}{arg\,min}
\begin{document}
\bibliographystyle{IEEEtran}

\title{
\vspace{-4mm}
\huge{Graph Coloring Based Pilot Allocation to Mitigate Pilot Contamination for Multi-Cell Massive MIMO Systems}}

\author{\IEEEauthorblockN{Xudong Zhu, Linglong Dai, and Zhaocheng Wang}



\vspace*{-8mm}

}
\maketitle

\begin{abstract}
A massive multiple-input multiple-output (MIMO) system, which utilizes a large number of base station (BS) antennas to serve a set of users, suffers from pilot contamination due to the inter-cell interference (ICI).
In this letter, a graph coloring based pilot allocation (GC-PA) scheme is proposed to mitigate pilot contamination for multi-cell massive MIMO systems.
Specifically, by exploiting the large-scale characteristics of fading channels, an interference graph is firstly constructed to describe the potential ICI relationship of all users.
Then, with the limited pilot resource, the proposed GC-PA scheme aims to mitigate the potential ICI by efficiently allocating pilots among users in the interference graph.
The performance gain of the proposed scheme is verified by simulations.
\end{abstract}

\begin{IEEEkeywords}
Massive multiple-input multiple-output (MIMO), pilot contamination, graph coloring, pilot allocation.
\vspace{-2mm}
\end{IEEEkeywords}
\IEEEpeerreviewmaketitle

\section{Introduction}
Massive multiple-input multiple-output (MIMO) has recently drawn extensive research interests due to its potential gains in spectrum efficiency and energy efficiency, whereby a base station (BS) equipped with a large number of antennas serves a set of users simultaneously \cite{noncooperative,dll}.
However, in practical multi-cell scenarios with unavoidable reuse of pilots in different cells, channel estimates obtained in a certain cell will be impaired by the reused pilots transmitted by users in other cells, which is referred as pilot contamination \cite{noncooperative}.
It is widely recognized that pilot contamination is the performance bottleneck of massive MIMO systems \cite{dll}.


How to mitigate pilot contamination has been extensively studied in the literature \cite{JS,blind1,emil}.
The greedy pilot allocation algorithm \cite{JS} can mitigate the pilot contamination by exploiting the statistical channel covariance information, but it suffers from high complexity.
A blind method based on subspace partitioning \cite{blind1} is able to reduce the inter-cell interference (ICI) when the channel vectors from different users are assumed to be orthogonal.
An adaptive pilot clustering method \cite{emil} based on coalitional game theory can mitigate the pilot contamination, whereby only a subset of the pilot resource is utilized in each cell, which leads to a slight loss in spectral efficiency.

The graph coloring algorithms (GCAs) developed from the graph theory, which is a powerful mathematical tool, have been widely utilized to reduce ICI in mobile communications \cite{Femtocell1,Femtocell2,D2D}.
For example, the available frequency bands are divided into several sub-bands and then allocated among femto-cells based on GCAs to strikes a balance between the ICI reduction and spatial frequency reuse \cite{Femtocell1,Femtocell2}.
A graph coloring based resource allocation scheme was proposed to reduce the ICI in the device-to-device (D2D) communication system by allocating different resource to different D2D pairs with severe ICI \cite{D2D}.
These existing works try to mitigate the ICI by allocating different resource in the time or frequency domain to different femto-cells or D2D pairs based on GCAs, but all of them lead to a significant loss of spectral efficiency.

In this letter, different from these existing works \cite{emil,Femtocell1,Femtocell2,D2D}, which are essentially based on frequency division, we consider a multi-cell multi-user massive MIMO system where all cells work in the same frequency band to achieve higher spectral efficiency, and a graph coloring based pilot allocation (GC-PA) scheme is proposed to reduce the pilot contamination.
Specifically, different from \cite{Femtocell1,Femtocell2} which construct the interference graph among femto-cells, we construct the interference graph based the graphic structure of the ICI relationship among users in all cells, whereby all users in the same cell are connected to each other due to the intra-cell interference, and users in different cells having the potential ICI stronger than a certain threshold are also connected.
After that, unlike \cite{Femtocell1,Femtocell2} which simply allocate different resource to different femto-cells, the proposed GC-PA scheme greedily allocates different pilots to the connected users in the interference graph under the constraints of limited pilot resources and non pilot reuse within one cell.
Simulation results verify the effectiveness and performance gain of the proposed GC-PA scheme in typical massive MIMO systems.

\section{System Model}

\begin{figure*}[t!]
\vspace{-5mm}
\center{\includegraphics[width=0.95\textwidth]{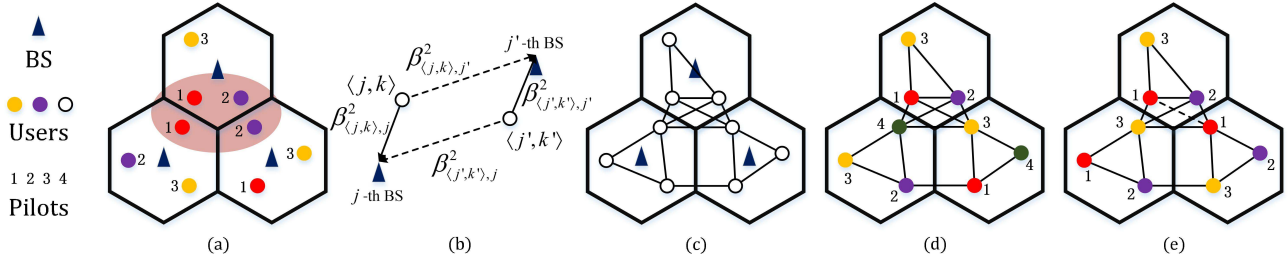}}
\vspace{-2mm}
\caption{A typical example:
(a) Classical random scheme: the available pilots will be randomly allocated to different users in each cell, which may cause severe ICI between users in adjacent cells, e.g., users in the shadow area;
(b) Potential ICI: the ratio of interference channel strength and effective channel strength;
(c) Construction of interference graph: an undirected graph can be constructed to describe the potential ICI relationship among all users;
(d) Conventional GCAs: Totally $C=4$ plots are required for $LK=9$ users;
(e) The proposed GC-PA scheme: only $K=3$ pilots are utilized for $LK=9$ users.}
\vspace{-5mm}
\label{forexample}
\end{figure*}

As shown in Fig. \ref{forexample} (a), we consider a multi-cell massive MIMO system composed of $L$ hexagonal cells, and each cell consists of a BS with $M$ antennas and $K$ ($K\ll M$) single-antenna users \cite{noncooperative}.
The channel vector $\mathbf{h}_{\langle j,k\rangle,i}\in\mathcal{C}^{M\times 1}$ from the user $\langle j,k \rangle$ ($k$-th user in the $j$-th cell) to the $i$-th BS can be modeled as
\begin{equation}
\mathbf{h}_{\langle j,k\rangle,i} = \mathbf{g}_{\langle j,k\rangle,i}\sqrt{\beta_{\langle j,k\rangle,i}},
\end{equation}
where $\beta_{\langle j,k\rangle,i}$ denotes the large-scale fading coefficient,
and $\mathbf{g}_{\langle j,k\rangle,i}\sim\mathcal{CN}(\mathbf{0},\mathbf{I}_M)$ denotes the small-scale fading vector.

According to \cite{noncooperative}-\cite{dll}, the pilots $\mathbf{\Phi}=[\boldsymbol{\phi}_1~\boldsymbol{\phi}_2~\cdots~\boldsymbol{\phi}_K]^T\in\mathcal{C}^{K\times\tau}$ with length $\tau$ ($\tau\geq K$) are orthogonal, i.e., $\mathbf{\Phi}\mathbf{\Phi}^H=\mathbf{I}_K$, and the pilot resource is reused in adjacent cells. 
The classical pilot allocation scheme randomly allocates the pilot $\boldsymbol{\phi}_{p_{\langle j,k\rangle}}$ to the user $\langle j,k \rangle$, and guarantees that every pilot will not be reused within one cell, i.e., $p_{\langle j,k\rangle}\neq p_{\langle j,k'\rangle}, \forall k\neq k'$ \cite{dll}.
According to \cite{noncooperative}, the uplink signal-to-interference-plus-noise-ratio (SINR) of the user $\langle j,k \rangle$ can be calculated as
\begin{eqnarray}\label{SINRUL}
\text{SINR}_{\langle j,k\rangle}^{\text{UL}}&=&\frac{|\mathbf{h}_{\langle j,k\rangle,j}^H\mathbf{h}_{\langle j,k\rangle,j}|^2}{\sum\limits_{\langle j',k' \rangle\in \mathcal{I}_{\langle j,k\rangle}}|\mathbf{h}_{\langle j',k'\rangle,j}^H\mathbf{h}_{\langle j',k'\rangle,j}|^2+\sigma_{\langle j,k\rangle}^2}\nonumber\\
&\rightarrow&\frac{\beta_{\langle j,k\rangle,j}^2}{\sum\limits_{\langle j',k' \rangle\in \mathcal{I}_{\langle j,k\rangle}}\beta_{\langle j',k'\rangle,j}^2}, \;\; M\rightarrow\infty,
\end{eqnarray}
where $\sigma_{\langle j,k\rangle}^2$ denotes the power of uncorrelated interference and noise which can be substantially reduced by increasing the number of BS antennas, $\sum_{\langle j',k' \rangle\in \mathcal{I}_{\langle j,k\rangle}}\beta_{\langle j',k'\rangle,j}^2$ denotes pilot contamination caused by pilot reuse, and $\mathcal{I}_{\langle j,k\rangle}$ denotes the set of users with the same pilot as user $\langle j,k \rangle$, which is defined as
$\mathcal{I}_{\langle j,k\rangle}=\big\{\langle j',k' \rangle: p_{\langle j',k'\rangle}=p_{\langle j,k\rangle}\big\}\setminus\big\{\langle j,k\rangle\big\}.$
Thus, the corresponding average uplink achievable rate of the user $\langle j,k \rangle$ can be calculated as
\begin{equation}\label{capacity}
    \text{C}_{\langle j,k\rangle}^{\text{UL}}=\big(1-\mu_0\big)\text{E}\Big\{\log_2(1+\text{SINR}_{\langle j,k\rangle}^{\text{UL}})\Big\},
\end{equation}
where $\mu_0$ evaluates the loss of spectral efficiency caused by the uplink pilot transmission, which is actually the ratio of the pilot sequence's length $\tau$ and the channel coherence time $\iota$ \cite{dll}, i.e., $\mu_0=\frac{\tau}{\iota}$.

It has been proved that the average uplink achievable rate is limited by pilot contamination \cite{noncooperative}.
For example, as shown in Fig. \ref{forexample} (a), the users in the shadow area suffer from severe pilot contamination due to the classical random pilot allocation scheme.
It should be pointed out that above definitions and derivations can be directly applied in practical cellular network with irregular geometry \cite{3GPP}.

\section{Graph Coloring based Pilot allocation}


\subsection{Construction of Interference Graph}
From Eq. (\ref{SINRUL}), it is clear that the ICI strength to a specific user $\langle j,k \rangle$ can be measured by $\sum_{\langle j',k' \rangle\in \mathcal{I}_{\langle j,k\rangle}}\beta_{\langle j',k'\rangle,j}^2$.
Thus, we define the metric $\eta_{\langle j,k\rangle,\langle j',k'\rangle}$ to measure the strength of potential ICI between user $\langle j,k\rangle$ and $\langle j',k'\rangle$ in different cells ($j\neq j'$) when they are allocated with the same pilot, i.e.,
\begin{align}\label{eta}
\eta_{\langle j,k\rangle,\langle j',k'\rangle}={\beta_{\langle j',k'\rangle,j}^2}/{\beta_{\langle j,k\rangle,j}^2}+{\beta_{\langle j,k\rangle,j'}^2}/{\beta_{\langle j',k'\rangle,j'}^2}.
\end{align}
An intuitive illustration of $\eta_{\langle j,k\rangle,\langle j',k'\rangle}$ is shown in Fig. \ref{forexample} (b), which indicates that $\eta$ is the ratio of the interference channel strength and the effective channel strength.
Larger $\eta_{\langle j,k\rangle,\langle j',k'\rangle}$ indicates more severe ICI will be introduced between $\langle j,k\rangle$ and $\langle j',k'\rangle$ when the same pilot is allocated to them.

In order to construct the interference graph to describe the potential ICI relationship among all users in the multi-cell massive MIMO system, a threshold $\gamma_{\text{th}}$ is introduced to determine whether two users in different cells can reuse the same pilot, i.e., $\eta_{\langle j,k\rangle,\langle j',k'\rangle}>\gamma_{\text{th}}$, where $\gamma_{\text{th}}$ will be discussed in detail in Section III-C.
Based on $\gamma_{\text{th}}$, we can generate a binary potential ICI matrix $\mathbf{B}=[b_{\langle j,k\rangle,\langle j',k'\rangle}]_{LK\times LK}$, where each entry $b_{\langle j,k\rangle,\langle j',k'\rangle}$ is defined as
\begin{equation}\label{B}
b_{\langle j,k\rangle,\langle j',k'\rangle}=\left\{
                \begin{array}{ll}
                  1, & j=j',~k\neq k',\\
                  1, & j\neq j',~\eta_{\langle j,k\rangle,\langle j',k'\rangle}> \gamma_{\text{th}}, \\
                  0, & \text{otherwise}.
                \end{array}
              \right.
\end{equation}
Intuitively, all users in the same cell are connected to each other, and two users in different cells are connected if their potential ICI strength is stronger than the threshold $\gamma_{\text{th}}$.

As illustrated in Fig.~\ref{forexample} (c), an interference graph can be constructed as an undirected graph $G=(\mathcal{V},\mathcal{E})$ based on the potential ICI matrix $\mathbf{B}$ from Eq. (\ref{B}), where the vertexes in set $\mathcal{V}$ denote all users, and the edges in set $\mathcal{E}$ denote potential ICIs among users.


\subsection{Proposed GC-PA Scheme}
GCAs developed in the graph theory is able to color the vertexes of a graph with the minimum number of colors, $C$, such that no two connected vertexes will share the same color \cite{Femtocell1}.
Equivalently, for the pilot allocation problem in massive MIMO systems, GCAs can ensure that any connected users will be allocated with different pilots to mitigate the potential pilot contamination.
Unfortunately, the minimum kinds of colors required by the conventional GCAs, i.e., $C$, is variable and unpredictable, since the interference graph consisted of moving users is dynamically changing due to the user mobility.
For example, as shown in Fig. \ref{forexample} (d), totally 4 kinds of colors are desirable to realize the graph coloring process, i.e., $C=4$.
However, in typical massive MIMO systems, only $K=3$ pilots are available to ensure orthogonal pilots used within one cell, and thus the conventional GCAs could not be applied.

\begin{algorithm}[htb]
\renewcommand{\algorithmicrequire}{\textbf{Input:}}
\renewcommand\algorithmicensure {\textbf{Output:} }
\caption{Proposed GC-PA Scheme}
\begin{algorithmic}[1]
\REQUIRE
~~~System parameters: $K$, $L$, and $\gamma_{\text{th}}$;\\
~~~~~~~\hspace{0.5mm}Large-scale fading coefficients: $\beta_{\langle j,k\rangle,i}$.\\
\ENSURE
Pilot allocation: $\{p_{\langle j,k\rangle}\}$. \\
\STATE $\;\;\;\;\{p_{\langle j,k\rangle}\}=0$, $\Omega=\varnothing$, calculate $b_{\langle j,k\rangle,\langle j',k'\rangle}$.
\WHILE {$\exists p_{\langle j,k\rangle}=0$}
\STATE $\langle j_0,k_0\rangle=\argmax\limits_{\langle j,k\rangle\notin\Omega}\big\{\sum\limits_{j\neq j'} b_{\langle j,k\rangle,\langle j',k'\rangle}\big\}$.
\STATE $\Lambda=\{1,2,\cdots,K\}\setminus\{p_{\langle j_0,k\rangle}:1\leq k\leq K\}$.
\STATE $\lambda_0=\argmin\limits_{\lambda\in\Lambda}\big\{\sum\limits_{p_{\langle j,k\rangle}=\lambda} b_{\langle j_0,k_0\rangle,\langle j,k\rangle} \big\}$.
\STATE $p_{\langle j_0,k_0\rangle}=\lambda_0$, $\Omega = \Omega\cup\langle j_0,k_0\rangle$.
\ENDWHILE
\end{algorithmic}
\end{algorithm}

In order to address this problem, the GC-PA scheme is proposed to mitigate the potential ICI with the constraint of predefined pilot resource for various user topology.
The proposed GC-PA scheme greedily allocates the available pilot with smallest ICI to the most important user in a sequential way, and the pseudocode is provided in {\bf Algorithm 1}.
Given the input information about cellular network parameters $L,K$, the threshold $\gamma_{\text{th}}$, and the large-scale fading coefficients $\beta_{\langle j,k\rangle,i}$, this algorithm outputs the pilot allocation information $p_{\langle j,k\rangle}$ for all users.
The physical meaning of {\bf Algorithm 1} can be explained step by step as follows:
\subsubsection{Initialization (step 1)}
The pilot allocation is firstly initialized, i.e., $\{p_{\langle j,k\rangle}\}=0$.
Then, the set $\Omega$ containing the users which have been allocated with their corresponding pilots is also initialized as a null set, i.e., $\Omega=\varnothing$.
Given $\beta_{\langle j,k\rangle,i}$, the binary potential ICI relationship $b_{\langle j,k\rangle,\langle j',k'\rangle}$ can be calculated based on Eq.~(\ref{B}).

\subsubsection{Loop Condition (step 2)}
The users will be allocated with their corresponding pilots in turns, and thus the loop condition is that there is at least one user who has not been allocated with any pilot, i.e., $\exists p_{\langle j,k\rangle}=0$.

\subsubsection{User Selection (step 3)}
Considering the fact that more connections in the interference graph lead to severer ICI, the user $\langle j_0,k_0\rangle$, who has the maximum connections in adjacent cells and has not been allocated with any pilot, will be selected.

\subsubsection{Available Pilot Set (step 4)}
To avoid the intra-cell interference, no pilot will be reused within the same cell \cite{noncooperative,dll}.
The pilots, which have not been allocated to users in the $j_0$-th cell, consists the available pilot set $\Lambda$ for user $\langle j_0,k_0\rangle$.

\subsubsection{Pilot Selection (step 5)}
In order to allocate different pilots to connected users to mitigate ICI as much as possible, the pilot with index $\lambda_0$, which has the minimal reused times among the connections of the selected user $\langle j_0,k_0\rangle$, will be selected from the available pilot set $\Lambda$ for user $\langle j_0,k_0\rangle$.

\subsubsection{Pilot Allocation (step 6)}
In each loop, the selected pilot with index $\lambda_0$ will be allocated to the selected user $\langle j_0,k_0\rangle$, and then this user will be added into the set $\Omega$.
The procedure from step $2\sim 6$ will be carried out for $KL$ times until all users are allocated with their corresponding pilots.

In contrast to the conventional GCAs which requires an unpredictable number of pilots, the proposed GC-PA scheme is able to solve the pilot allocation problem with restricted $K$ pilots.
For instance as shown in Fig. \ref{forexample} (e), total $K=3$ pilots are utilized to realize pilot allocation with significantly reduced ICI by the proposed GC-PA scheme.

\subsection{Analysis of Threshold $\gamma_{\text{th}}$}
As the threshold $\gamma_{\text{th}}$ is used to be compared with the proposed heuristic metric $\eta_{\langle j,k\rangle,\langle j',k'\rangle}$, it is a key parameter for the construction of the interference graph and the performance of the proposed GC-PA scheme.

From Eq.~(\ref{B}), we can find that any threshold chosen from $\big(-\infty,\eta_{\text{min}}\big)$ leads to the same interference graph with the threshold $\gamma_{\text{th}}=\eta_{\min}$, where $\eta_{\text{min}}$ denotes the minimum value of $\eta_{\langle j,k\rangle,\langle j',k'\rangle}$ based on Eq.~(\ref{eta}).
Similarly, the upper bound of the threshold $\gamma_{\text{th}}$ can be set as $\eta_{\max}$, so we have
\begin{align}
\gamma_{\text{th}}\in\big[\eta_{\min},\eta_{\max}\big].
\end{align}

In one extreme case, when the threshold $\gamma_{\text{th}}$ is set to $\eta_{\max}$, only users within the same cell will be connected to each other.
In this case, both the GCAs and the proposed GC-PA scheme degrade to the classical random pilot allocation scheme, since no ICI information is exploited.
By decreasing the threshold $\gamma_{\text{th}}$, more users in different cells will be connected to reflect the potential ICI relationship.
In this case, as discussed above, the GCAs are able to eliminate the ICI among the connected users with an increasing number of pilots $C$, while the proposed GC-PA scheme can also mitigate most ICI with the restricted $K$ pilots, where $K<C$.
In the other extreme case, when the threshold $\gamma_{\text{th}}$ is set to $\eta_{\min}$, all $KL$ users will be connected to each other.
It is evident that total $C=KL$ pilots are required by the GCAs in this case, while the proposed GC-PA scheme degrades to the classical random pilot allocation scheme.

To obtain a near-optimal threshold $\gamma_{\text{th}}$ in practical systems, we propose a simple yet effective method called iterative grid search (IGS) as
\begin{align}
\gamma_{\text{th}}=f_{\text{IGS}}\Big([\eta_{\min},\eta_{\max}], N, T\Big),
\end{align}
where $N$ denotes the number of grids in each search step, and $T$ denotes the number of iterations.
Specifically, the interval $[\eta_{\min},\eta_{\max}]$ is uniformly sampled by $N$ points in the first iteration, i.e., $\{\gamma_i^{(1)}:1\leq i\leq N\}$, $\gamma_0^{(1)}=\eta_{\min}$, $\gamma_{i+1}^{(1)}=\gamma_i^{(1)}+\Delta^{(1)}$, and $\Delta^{(1)}=\frac{\eta_{\max}-\eta_{\min}}{N-1}$.
By denoting $\gamma_{\max}^{(1)}$ as the one out of $\{\gamma_i^{(1)}\}$ that can achieve the highest average uplink SINR, a sub interval $[\gamma_{\max}^{(1)}-\Delta^{(1)}/2,\gamma_{\max}^{(1)}+\Delta^{(1)}/2]$ can be obtained after the first iteration, which will be further sampled in the next iteration to obtain a smaller range $[\gamma_{\max}^{(2)}-\Delta^{(2)}/2,\gamma_{\max}^{(2)}+\Delta^{(2)}/2]$.
This procedure is carried out in an iterative way for $T$ times, and finally a near-optimal threshold $\gamma_{\text{th}}=\gamma_{\max}^{(T)}$ can be obtained.

\subsection{Discussion of Implementation}
In Section III-A, the interference graph is constructed according to the large-scale fading coefficients $\beta_{\langle j,k\rangle,i}$, which actually denotes the channel strength between the users and the BSs.

Compared with the existing works \cite{JS,blind1} which assume the statistical channel covariance can be accurately acquired, the proposed GC-PA scheme only requires large-scale fading coefficients, which changes slowly and can be easily tracked with low complexity \cite{noncooperative,dll,3GPP}.
For example, in practical mobile cellular networks of the long term evolution (LTE) systems \cite{3GPP}, an user firstly captures the cell-specific broadcasting signals to measure the channel conditions to available BSs, and then selects the BS with the best channel condition.
After that, this user will continue to track the channel conditions to available BSs, and thus the handover among different cells can be easily realized.

As addressed in LTE systems \cite{3GPP}, the BSs in adjacent cells are connected to each other through X2 interface to share and exchange information.
Moreover, the coordinated BSs are all connected to a mobility management entity (MME) through S1 interface, which has a powerful computing capability.
Thus, the MME is able to collect the large-scale fading coefficients from coordinated BSs, and operate the proposed GC-PA scheme to provide a near-optimal pilot allocation pattern.

According to {\bf Algorithm 1}, the complexity $\mathcal{O}(K^2 L^2)$ is required to operate the proposed GC-PA scheme based on a specific threshold $\gamma_{\text{th}}$.
According to the IGS procedure, {\bf Algorithm 1} will be carried out for $NT$ times to achieve the near-optimal threshold $\gamma_{\text{th}}$, and thus the total complexity is $\mathcal{O}(NTK^2 L^2)$, which is acceptable for the powerful MME \cite{3GPP}.

\section{Numerical Results}
\begin{figure*}[!t]
\centering
\vspace{-10mm}
\subfigure[] {\includegraphics[height=1.8in,width=2.3in]{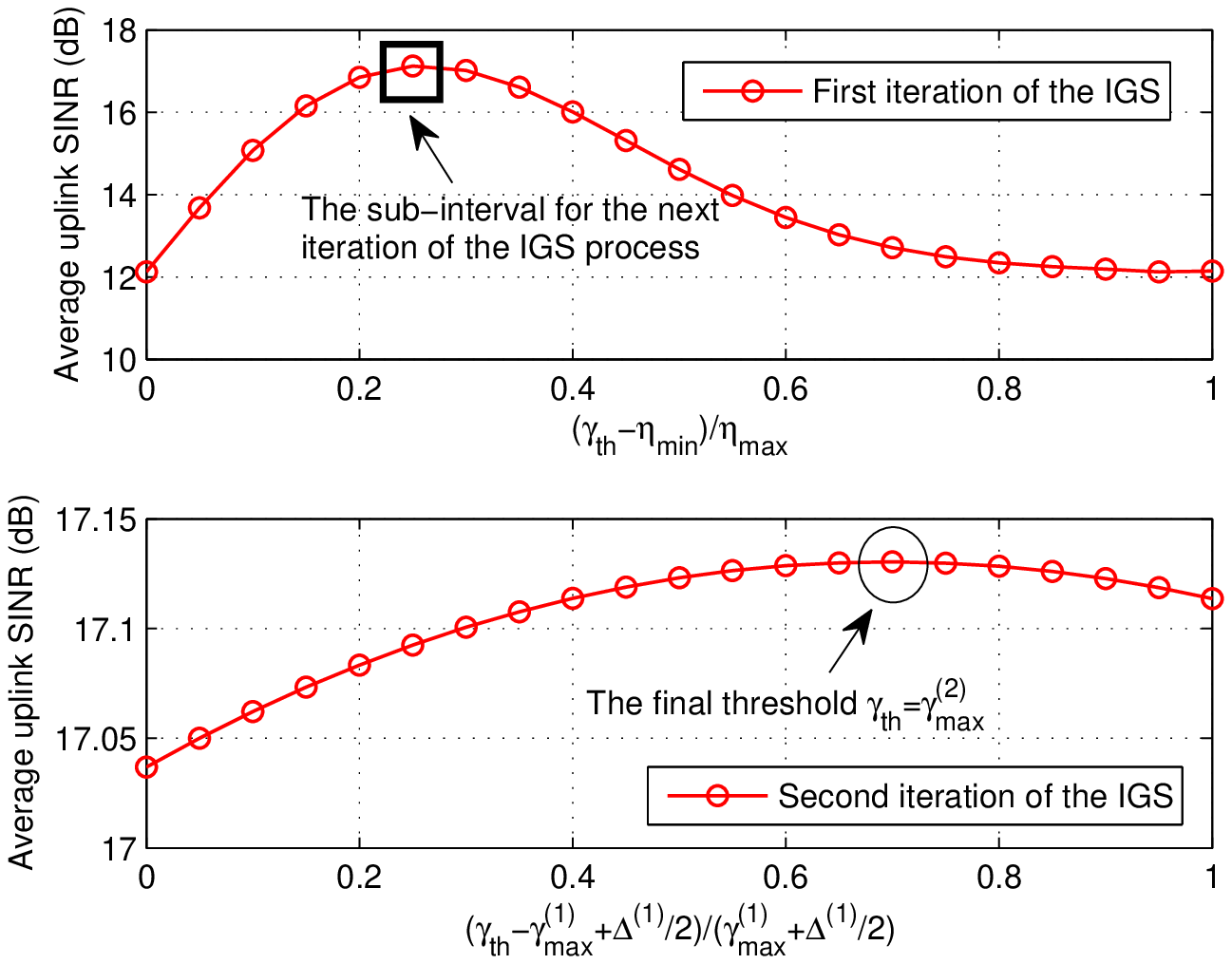}}
\subfigure[] {\includegraphics[height=1.8in,width=2.3in]{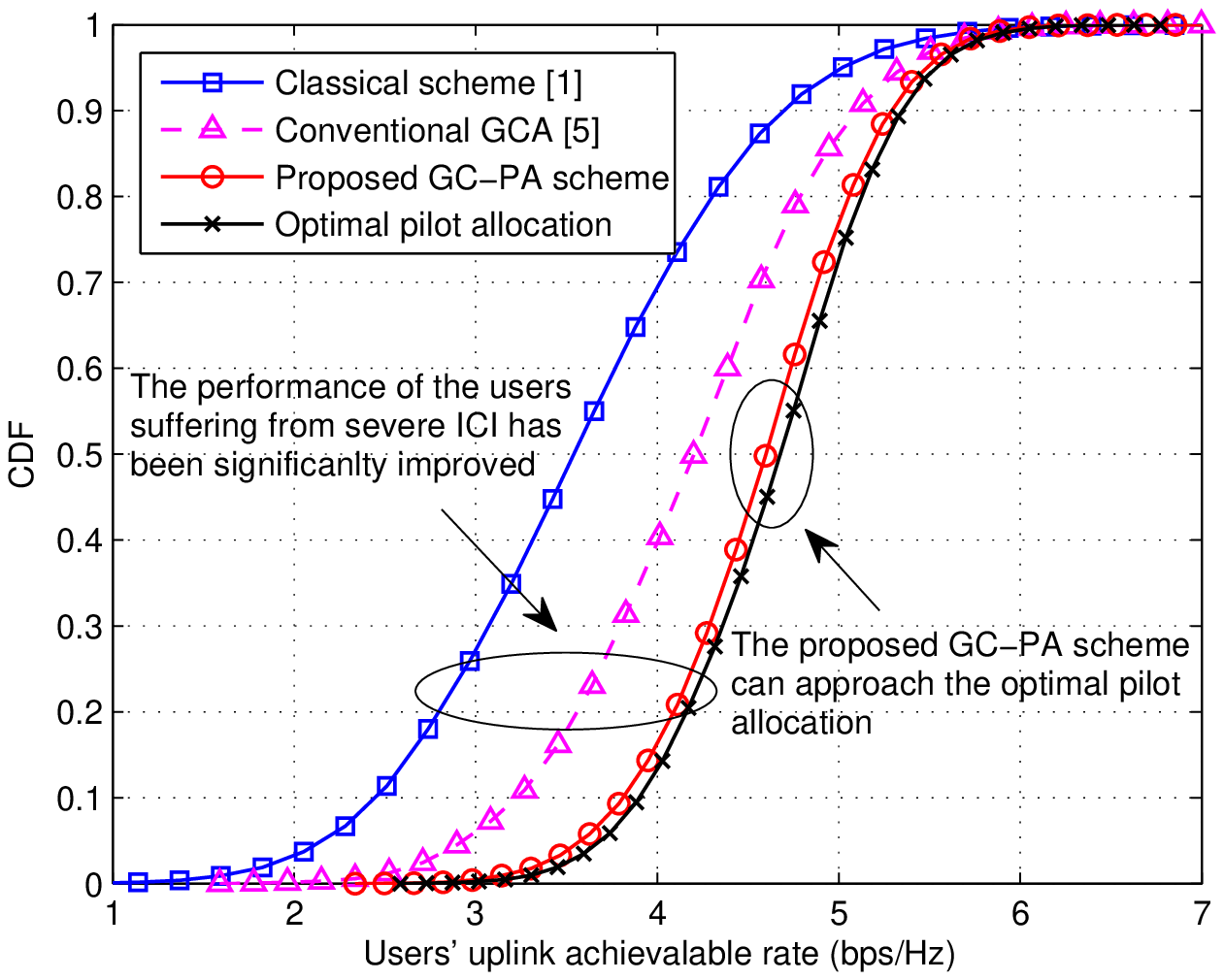}}
\subfigure[] {\includegraphics[height=1.8in,width=2.3in]{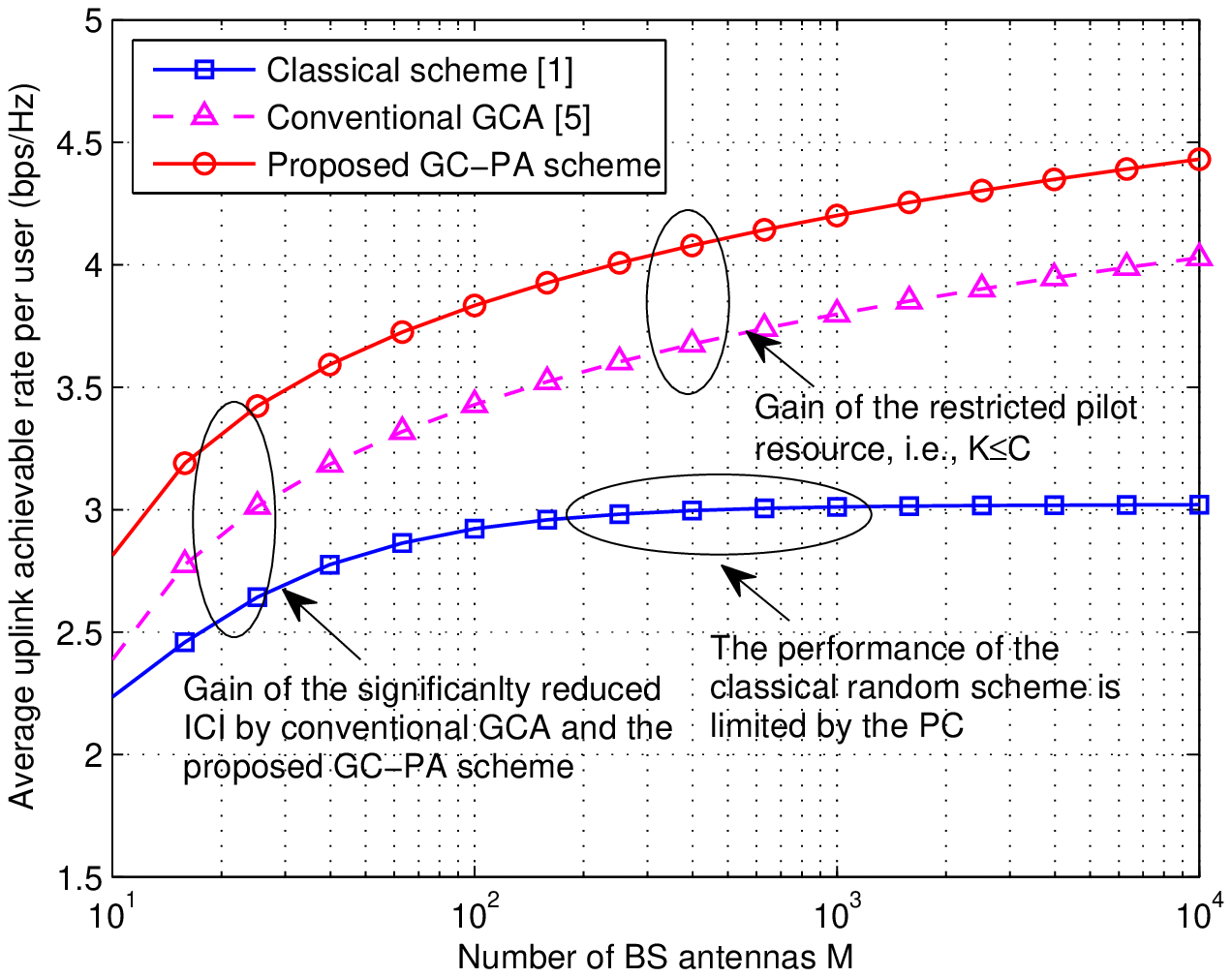}}
\vspace{-2mm}
\caption{Simulation results:
(a) The IGS procedure to obtain the near-optimal threshold $\gamma_{\text{th}}$ with $L=7$, $K=8$, $M=128$, $N=20$, and $T=2$;
(b) The CDF of the users' uplink achievable rate with $L=4$, $K=4$, and $M=128$;
(c) The average uplink achievable rate per user against the number of BS antennas $M$ with $L=7$ and $K=8$.
}
\label{FigSimulation}
\vspace{-6mm}
\end{figure*}

In this section, we consider a typical hexagonal cellular network with $L$ cells, where each cell has $K$ users with single-antenna and a BS with $M$ antennas \cite{noncooperative,dll,JS,blind1}, and the default values of the various parameters are listed in Table \ref{parameters}.
As addressed in \cite{noncooperative}, the large-scale fading coefficient $\beta_{\langle j,k\rangle,i}$ can be modeled as $\beta_{\langle j,k\rangle,i}={z_{\langle j,k\rangle,i}}/{(r_{\langle j,k\rangle,i}/R)^{\alpha}}$,
where $z_{\langle j,k\rangle,i}$ denotes the shadow fading and possesses a log-normal distribution, i.e., $10\log_{10}(z_{\langle j,k\rangle,i})\sim\mathcal{CN}(0,\sigma_{\text{shadow}})$, and $r_{\langle j,k\rangle,i}$ is the distance from the user $\langle j,k\rangle$ to the $i$-th BS.

\begin{table}
\vspace{2mm}
\caption{Simulation Parameters}
\centering
\label{parameters}
\vspace{-2mm}
\begin{tabular}{l|l}
  \hline\hline
  Number of cells $L$ & 4, 7 \\\hline
  Number of BS antennas $M$ & $10\leq M\leq 10000$ \\\hline
  Number of users in each cell $K$ & 4, 8 \\\hline
  Cell radius $R$ & 500 m\\\hline
  Path loss exponent $\alpha$ & 3 \\\hline
  Log normal shadowing fading $\sigma_{\text{shadow}}$ & 8 dB \\\hline
  Loss of spectral efficiency $\mu_0$ & $\mu_0=0.2$\\\hline
  IGS parameters & $N=20,T=2$\\\hline
  \hline
\end{tabular}
\vspace{-5mm}
\end{table}

Fig. \ref{FigSimulation} (a) shows the IGS procedure to obtain the near-optimal threshold $\gamma_{\text{th}}$ when $L=7$, $K=8$, $M=128$, and $N=20$ are considered as an example, and we only select $T$ as $T=2$.
The X-axis denotes the normalized threshold interval, and the Y-axis denotes the average uplink SINR of the proposed GC-PA scheme with the thresholds $\{\gamma_i^{(t)}\}$, which are the uniformly sampled points in the $t$-th iteration.
It is evident that the best sub-interval is selected after the first iteration of the IGS procedure, and a near-optimal threshold $\gamma_{\text{th}}=\gamma_{\max}^{(2)}$ can be obtained after the second iteration.

Fig. \ref{FigSimulation} (b) plots the cumulative distribution function (CDF) curve of the users' uplink achievable
rate with $L=4$, $K=4$, and $M=128$.
The classical scheme randomly allocates the available pilots to all users \cite{noncooperative}, the conventional GCA is able to eliminate all potential ICI by consuming unrestricted number of pilots $C$ \cite{Femtocell1,D2D}, while the optimal pilot allocation is obtained through exhaustive search among all possible cases as much as $(K!)^{L-1}=(4!)^3=13824$.
The curve of the conventional GCA is denoted by a dotted line, since they could not be applied in practical systems with predefined and restricted pilot resource \cite{noncooperative,dll,3GPP}.
It should be pointed out that, $\mu_C=\mu_0\frac{C}{K}$ instead of $\mu_0$, is utilized to calculate the users' uplink achievable rate of the conventional GCA.
From Fig. \ref{FigSimulation} (b), it is evident that the conventional GCA outperforms the classical scheme by about 0.6 bps/Hz, and another gain of about 0.4 bps/Hz can be achieved by the proposed GC-PA scheme.
Moreover, the performance gap between the proposed GC-PA scheme and the optimal PA is only about 0.1 bps/Hz.

Fig.~\ref{FigSimulation} (c) presents the average uplink achievable rate per user against the number of BS antennas $M$ with $L=7$ and $K=8$.
The curve of the optimal pilot allocation is not plotted here since its computational complexity is too high to be realized, i.e., $(K!)^{L-1}=(8!)^6\approx 4.3\times 10^{27}$.
It is evident that the proposed GC-PA scheme outperforms the classical scheme by about 1.0 bps/Hz when the typical number of BS antennas $M=128$ is considered.
Moreover, it is noteworthy that the performance of the proposed GC-PA scheme can be continually improved by increasing the number of BS antennas $M$, while that of the classical scheme becomes saturated fast.

\section{Conclusions}
In this letter, we have proposed a pilot allocation scheme based on graph coloring to mitigate pilot contamination for multi-cell massive MIMO systems.
Firstly, an interference graph is constructed to describe the potential ICI relationship among all users.
After that, the GC-PA scheme is proposed to eliminate most potential ICI in the interference graph.

\end{document}